\documentclass[12pt,preprint]{aastex}

\def\deg{{$^{\circ}$}}
\def\bd17{\mbox{BD +17\deg 3248}}
\def\cs22{\mbox{CS 22892-052}}
\def\gtaprx{ \mathrel{  \vcenter{
                        \offinterlineskip \hbox{$>$}
                        \kern 0.3ex \hbox{$\sim$}    } } }

\def\etal{\mbox{et al.}}
\received{}
\accepted{}
\journalid{}{}
\articleid{}{}
\paperid{}
\cpright{AMES LAW}{2006}
\ccc{}

\slugcomment{{\it NASA LAW}, February 14-16, 2006, UNLV, Las Vegas }

\shorttitle{The Short Title of Your Manuscript Goes Here}
\shortauthors{F1. M1. Name1 {\em et al}}

\begin{document}

\title{Nucleosynthesis: Stellar and Solar Abundances and Atomic Data}

\author{
John J. Cowan,\altaffilmark{1}
James E. Lawler,\altaffilmark{2}
Christopher Sneden,\altaffilmark{3}
E. A. Den Hartog,\altaffilmark{2}\\
and
Jason Collier\altaffilmark{1}
}

\altaffiltext{1}{Homer L. Dodge Department of Physics and Astronomy,
University of Oklahoma, Norman, OK 73019; cowan@nhn.ou.edu, collier@nhn.ou.edu}
\altaffiltext{2}{Department of Physics, University of Wisconsin, Madison,
Madison, WI 53706; jelawler@wisc.edu, eadenhar@wisc.edu}
\altaffiltext{3}{Department of Astronomy and
McDonald Observatory, University of Texas,
Austin, Texas 78712; chris@verdi.as.utexas.edu}

\begin{abstract}
Abundance observations  indicate the presence of
often surprisingly large amounts of 
neutron capture ({\it i.e.}, $s$- and $r$-process) elements in old Galactic halo
and globular cluster stars.
These observations provide insight into the nature of
the earliest generations of stars in the Galaxy -- the progenitors of the
halo stars --
responsible
for neutron-capture synthesis. Comparisons of abundance trends can be
used to understand the chemical evolution of the Galaxy and the nature
of heavy element nucleosynthesis. In addition age determinations, based upon
long-lived radioactive nuclei abundances, can now be obtained.
These stellar abundance determinations  depend critically upon atomic
data. 
Improved laboratory transition probabilities have
been recently obtained for a number of elements.
These new gf values have been used to greatly refine the abundances 
of neutron-capture elemental abundances in the solar photosphere and 
in very metal-poor Galactic halo stars.
The newly determined stellar abundances
are surprisingly consistent with a (relative)  Solar System
$r$-process pattern,
and are also consistent with abundance predictions
expected from such neutron-capture nucleosynthesis.
   
\end{abstract}



\section {Stellar Abundances and New Atomic Data}

Abundance studies of heavy elements and isotopes in Galactic halo stars
are providing important clues and insights into the  nature of, and 
sites for, nucleosynthesis
early in the history of the Galaxy 
(Sneden \& Cowan 2003; Cowan \& Thielemann 2004; Cowan \& Sneden 2006). 
Observations of heavy neutron-capture 
elements ({\it i.e.}, those produced in the slow ($s$) or 
rapid ($r$)-process) in both old (low iron, or metallicity) stars and
younger (more metal-rich) stars 
are also providing clear indications of the nature
of the Galactic chemical evolution. In addition the detection of the 
long-lived radioactive 
$n$-capture elements, such as thorium and uranium, are allowing 
direct age determinations for the oldest stars in the Galaxy - thus, providing
lower limits on both the age of the Galaxy and the Universe.

\begin{figure}[ht]
\centering
\includegraphics[angle=0,width=3.20in]{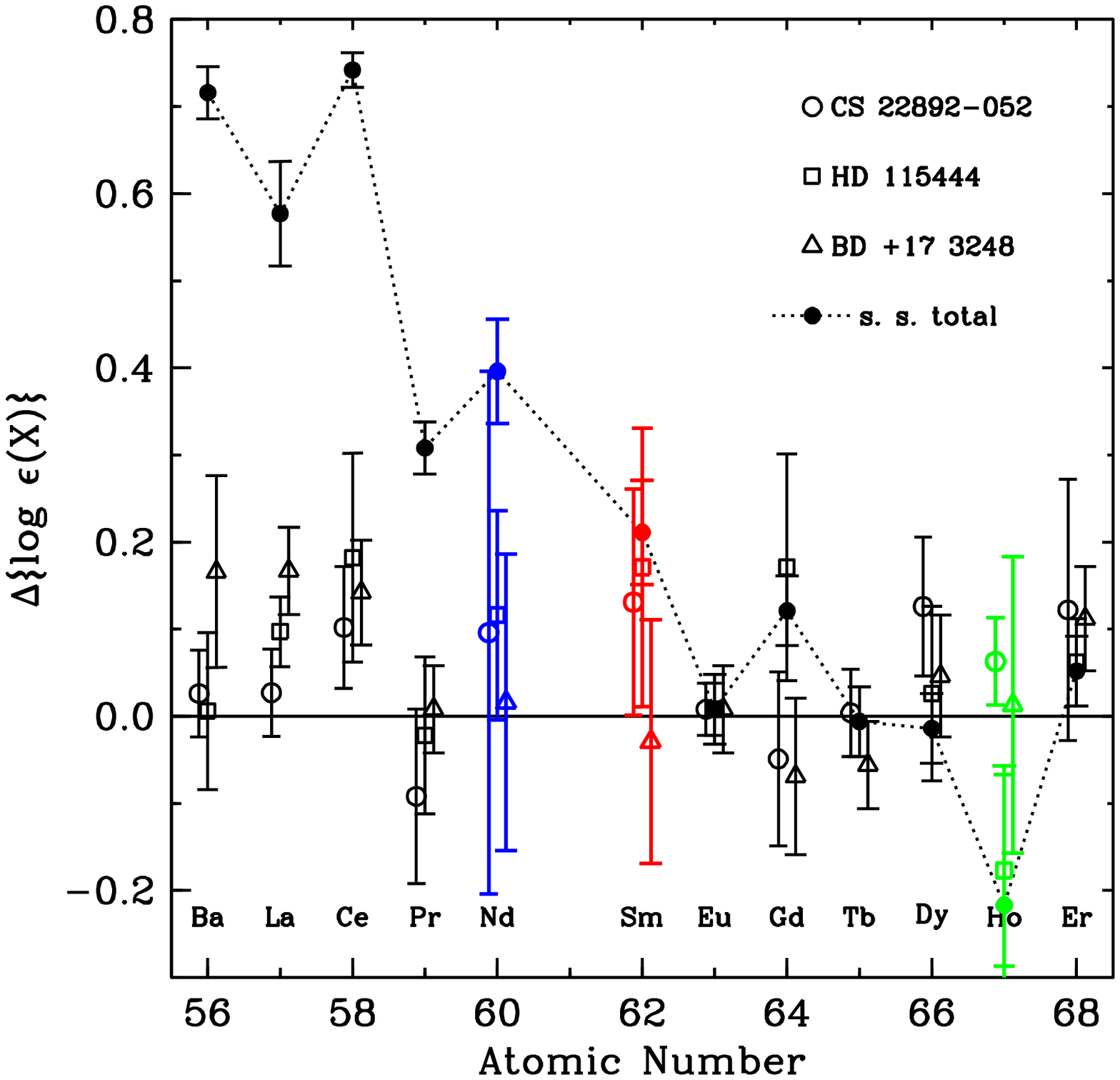}
\includegraphics[angle=0,width=3.20in]{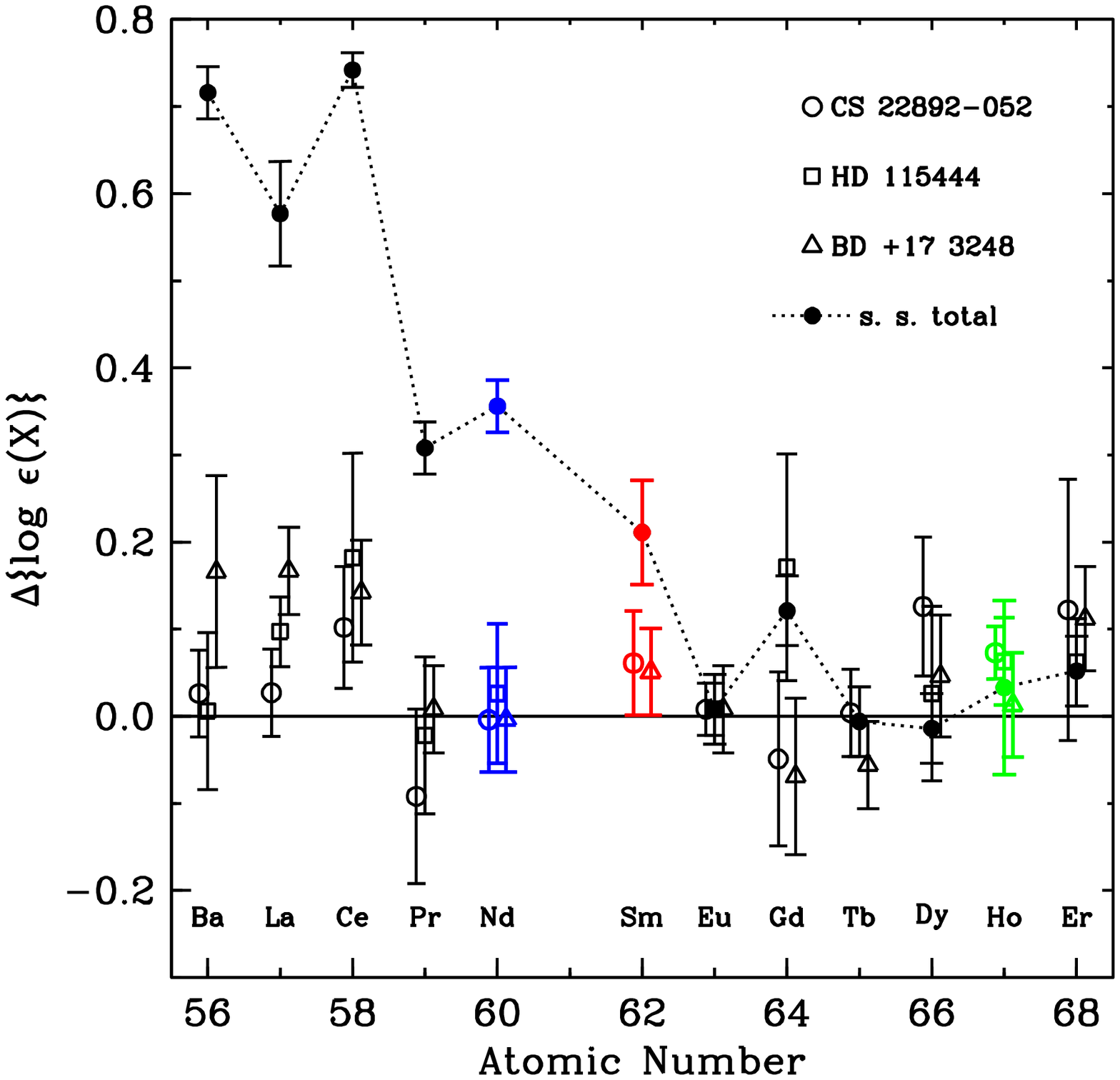}
\vskip -0.9in
\caption{(left) Abundance values (scaled at the element Eu) 
for selected elements in the stars \cs22 (open circles), HD 115444 (squares),
\bd17 (triangles) and the Sun (filled circles),  
based upon previously published atomic 
data.
(right) Newly derived abundance values based upon recent experimental lab data 
(after Lawler \etal\ 2006).
\label{fig1}}
\end{figure}

All of these abundance determinations and studies depend critically
upon atomic data, particularly transition probabilities.  
During the past few years there have been dramatic improvements 
in the experimental determinations of these transition probabilities 
for a number of rare earth elements:  
these include 
improved laboratory values for the elements 
Ce (Palmeri \etal\ 2000), 
Nd (Den Hartog \etal\ 2003), 
Ho (Lawler \etal\ 2004), 
Pt (Den Hartog \etal\ 2005),  
and Sm (Lawler \etal\ 2006).  
New values for the element Gd have also recently been obtained (Den Hartog
\etal\ 2006), but Gd abundances in the current paper do not reflect these
new data.
These improved data sets have been employed to 
determine elemental abundances in three metal-poor Galactic halo stars.
In addition  
new (refined) solar 
photospheric abundances have been obtained 
for Nd, Ho and Sm - this was not possible, however,    
for Pt. 
We show in Figure \ref{fig1} 
the abundances from Ba-Er (normalized at the $r$-process element Eu) 
in \cs22 (Sneden \etal\ 2003), \bd17 (Cowan \etal\ 2002), HD 115444 
(Westin \etal\ 2000) and the Sun (Lodders 2003). We focus 
in the left panel on abundance determinations for Nd, Sm and Ho in these
four stars based upon published atomic data. The abundances for those same
three elements, employing the newer laboratory atomic data,  is shown in 
the right panel of this figure. It is clear,  that as a result of 
employing the new atomic data, 
the star-to-star scatter among the three metal-poor Galactic halo stars is
markedly reduced, and there is good agreement
between the elemental values in these stars 
and the solar system $r$-process value (indicated 
by the horizontal  line).

\section {Discussion}

\begin{figure}[ht]
\centering
\includegraphics[angle=0,width=3.50in]{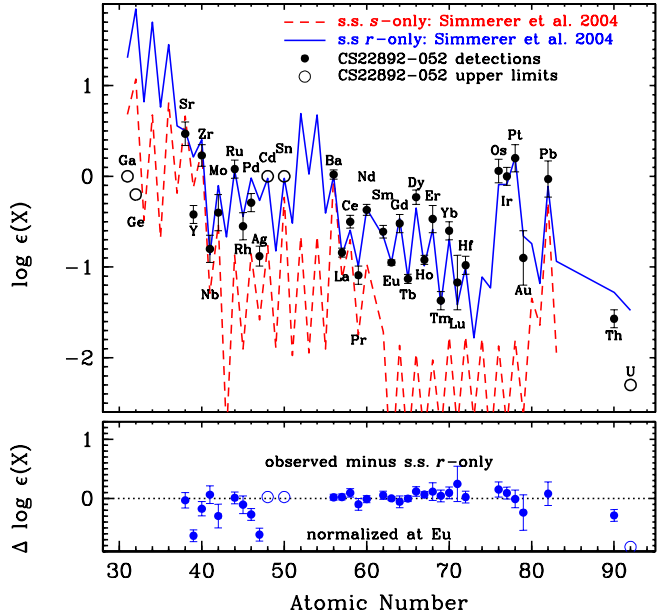}
\vskip -0.9in
\caption{(top) Abundance values for CS 22892-052  
compared to a scaled solar system $s$-process (dashed line) and
$r$-process distribution 
(solid line, Simmerer et al. 2004). (bottom) Differences between 
observed abundances in \cs22 and the scaled solar system r-process
abundances (after Cowan \& Sneden 2006).\label{fig2}}
\end{figure}

Employing the new atomic data, we have updated several of 
the abundance values from Sneden \etal\  (2003) for 
\cs22.  
We show those values in 
Figure \ref{fig2}  (top panel) compared to an $s$-only (dashed line) and 
an $r$-only (solid line) solar system elemental abundance curve.  
These solar system elemental abundance curves 
from Simmerer \etal\ (2004) are the sums of individual isotopic 
contributions from the $s$ and $r$-process  
and are tabulated in  
table~\ref{tbl1}. 
Using the so called ``classical model'' approximation 
in conjunction with measured neutron capture cross-sections, 
the individual $s$-process contributions are first determined.  
Subtracting these $s$-process isotopic contributions from the 
total solar
abundances  predicts  the residual $r$-process contributions.
These individual $s$- and $r$-process isotopic solar system  
abundances (based upon the Si = 10$^6$ scale) that are listed in table 1,   
are derived from  the work of K\"appeler \etal\ 1989, Wisshak \etal\ 
1998 and O'Brien \etal\ 2003, 
and can be used for future isotopic studies.

It is clear in both the top and bottom panels of Figure \ref{fig2} 
that the abundances of the stable elements Ba and above 
are consistent with the scaled solar system elemental $r$-process distribution.
This agreement has been seen in other $r$-process-rich stars and 
strongly suggests that the $r$-process is robust over the history of the
Galaxy. It further demonstrates that early in the history of the 
Galaxy, all of the $n$-capture elements were synthesized in the $r$-process,
and not the $s$-process. 
Additional comparisons in the figure, however, 
suggest the lighter elements do not fall
on this same curve that fits the heavier $n$-capture elements.
The upper limit on Ge,  for example, falls far below the scaled 
solar system $r$-process curve. 
Recent analyses of this element
indicates its abundance is tied to the iron level 
in metal-poor halo stars (Cowan \etal\ 2005),   
and thus suggests Ge synthesis in some 
type of
charged-particle reactions,
or other primary process,
in massive stars and supernovae
in the early Galaxy (see discussion in Cowan \& Sneden 2006). 
We also  note that the abundances of the $n$-capture elements from Z = 40--50 
in \cs22, in general, fall below the solar $r$-process curve. This has been 
interpreted as suggesting perhaps a second $r$-process synthesis site in 
nature (see discussion in Kratz \etal\ 2006 and Cowan \& Sneden 2006).
Interestingly,  new abundances studies of the star HD~221170 show 
better agreement with the scaled solar system curve for the elements
from Z = 40-50 than \cs22 does (Ivans \etal\ 2006). 

In spite of the good overall match between the scaled-solar $r$-process 
abundances and those of very metal-poor stars, some small deviations are 
becoming apparent, 
as shown in
Figures 1 and 2. 
The increasingly
accurate stellar abundance determinations, resulting in large 
part from the more accurate laboratory atomic data, 
are  helping to constrain,   
and ultimately could predict,  the actual values of the solar system 
$r$-process abundances.
Thus,  the current 
differences might suggest that some of those 
solar system $r$-process elemental predictions 
need to be reassessed. Recall that the elemental $r$-process curve
is based upon the isotopic $r$-process residuals, 
resulting from the subtraction of
the $s$-process isotopic contributions from the total solar system
abundances. Thus, a reanalysis of some of the $s$-process contributions,
and new neutron-capture cross sections measurements, might be in order. 
\section{Conclusions}

Abundance observations of metal-poor (very old) Galactic halo stars 
indicate the presence of $n$-capture elements. 
New laboratory  atomic data is dramatically reducing the stellar 
abundance uncertainties in these stars 
and increasingly improving the  
Solar System abundances. 
Recently determined abundances, based upon the new laboratory
data, 
indicate that the elemental  patterns in
the old halo stars are  consistent with each other and 
with a scaled solar system $r$-process 
abundance distribution  --  demonstrating that  
all of these elements (even $s$-process ones like Ba) 
were synthesized in the $r$-process early in the history of the Galaxy. 
These more accurate abundance determinations,  based upon more precise 
laboratory  atomic data, might be employed to 
constrain predictions for the solar system $r$-process abundances.
In the future additional (and improved) lab data  will 
have implications for a number of synthesis studies, 
including 
providing evidence regarding the possibility of two astrophysical
$r$-process sites. Such new experimental data 
will also help in understanding the Galactic chemical evolution of the 
elements, and could have an impact on chronometric 
age estimates of the Galaxy and the Universe. 
New stellar and atomic data in both the UV and IR wavelength regimes
may also allow the detection of 
never before seen elements in these stars, and 
will aid in new determinations of isotopic abundance mixtures  
for elements such as Sm.

\acknowledgments
We thank our colleagues for valuable insights and contributions.
This work has been supported in part by NSF
grants AST 03-07279 (J.J.C.),  
AST 05-06324 (J.E.L.),
AST 03-07495 (C.S.),
and by STScI.


\begin{center}
\begin{deluxetable}{ccccc|cccccc}
\tabletypesize{\scriptsize}
\tablenum{1}
\tablewidth{0pt}
\tablecaption{$s$- AND $r$-PROCESS ISOTOPIC SOLAR  SYSTEM 
ABUNDANCES\label{tbl1}}
\tablecolumns{12}
\tablehead{
\colhead{Element}        &
\colhead{Z}            &
\colhead{Isotope}        &
\colhead{N[s]}        &
\colhead{N[r]}          &
\colhead{Element}        &
\colhead{Z}            &
\colhead{Isotope}        &
\colhead{N[s]}        &
\colhead{N[r]}          &
 
}         
\startdata
Ga	&	31	&	69	&	11.137	&	11.600	&	Pd	&	46	&	104	&	0.165	&	0.000	\\
	&		&	71	&	10.413	&	4.700	&		&		&	105	&	0.040	&	0.269	\\
Ge	&	32	&	70	&	15.000	&	0.000	&		&		&	106	&	0.186	&	0.193	\\
	&		&	72	&	18.323	&	14.000	&		&		&	108	&	0.226	&	0.145	\\
	&		&	73	&	3.531	&	5.670	&		&		&	110	&	0.000	&	0.163	\\
	&		&	74	&	15.733	&	27.300	&	Ag	&	47	&	107	&	0.058	&	0.239	\\
	&		&	76	&	0.000	&	9.200	&		&		&	109	&	0.059	&	0.196	\\
As	&	33	&	75	&	1.456	&	5.330	&	Cd	&	48	&	110	&	0.178	&	0.000	\\
Se	&	34	&	76	&	4.656	&	0.000	&		&		&	111	&	0.042	&	0.165	\\
	&		&	77	&	1.679	&	3.040	&		&		&	112	&	0.198	&	0.196	\\
	&		&	78	&	7.402	&	7.210	&		&		&	113	&	0.060	&	0.139	\\
	&		&	80	&	7.446	&	24.300	&		&		&	114	&	0.287	&	0.140	\\
	&		&	82	&	0.000	&	5.710	&		&		&	116	&	0.000	&	0.121	\\
Br	&	35	&	79	&	0.450	&	0.000	&	In	&	49	&	115	&	0.057	&	0.121	\\
	&		&	81	&	0.479	&	4.640	&	Sn	&	50	&	116	&	0.511	&	0.000	\\
Kr	&	36	&	80	&	1.021	&	0.000	&		&		&	117	&	0.146	&	0.128	\\
	&		&	82	&	6.207	&	0.000	&		&		&	118	&	0.706	&	0.137	\\
	&		&	83	&	1.989	&	3.750	&		&		&	119	&	0.155	&	0.065	\\
	&		&	84	&	10.575	&	18.000	&		&		&	120	&	1.099	&	0.078	\\
	&		&	86	&	9.481	&	0.930	&		&		&	122	&	0.000	&	0.154	\\
Rb	&	37	&	85	&	0.690	&	2.790	&		&		&	124	&	0.000	&	0.199	\\
	&		&	87	&	2.214	&	0.100	&	Sb	&	51	&	121	&	0.047	&	0.113	\\
Sr	&	38	&	86	&	2.111	&	0.000	&		&		&	123	&	0.000	&	0.132	\\
	&		&	87	&	1.443	&	0.000	&	Te	&	52	&	122	&	0.133	&	0.000	\\
	&		&	88	&	16.986	&	2.550	&		&		&	123	&	0.047	&	0.000	\\
Y	&	39	&	89	&	3.344	&	1.310	&		&		&	124	&	0.248	&	0.000	\\
Zr	&	40	&	90	&	4.529	&	0.990	&		&		&	125	&	0.088	&	0.267	\\
	&		&	91	&	1.158	&	0.040	&		&		&	126	&	0.450	&	0.525	\\
	&		&	92	&	1.289	&	0.540	&		&		&	128	&	0.000	&	1.526	\\
	&		&	94	&	1.687	&	0.170	&		&		&	130	&	0.000	&	1.634	\\
	&		&	96	&	0.000	&	0.300	&	I	&	53	&	127	&	0.050	&	0.851	\\
Nb	&	41	&	95	&	0.229	&	0.110	&	Xe	&	54	&	128	&	0.126	&	0.000	\\
Mo	&	42	&	95	&	0.189	&	0.213	&		&		&	129	&	0.066	&	1.240	\\
	&		&	96	&	0.475	&	0.000	&		&		&	130	&	0.199	&	0.000	\\
	&		&	97	&	0.156	&	0.087	&		&		&	131	&	0.087	&	0.954	\\
	&		&	98	&	0.514	&	0.093	&		&		&	132	&	0.498	&	0.800	\\
	&		&	100	&	0.000	&	0.242	&		&		&	134	&	0.000	&	0.449	\\
Tc	&	43	&	99	&	0.006	&	0.172	&		&		&	136	&	0.000	&	0.373	\\
Ru	&	44	&	99	&	0.049	&	0.000	&	Cs	&	55	&	133	&	0.056	&	0.315	\\
	&		&	100	&	0.242	&	0.000	&	Ba	&	56	&	134	&	0.178	&	0.000	\\
	&		&	101	&	0.050	&	0.266	&		&		&	135	&	0.068	&	0.298	\\
	&		&	102	&	0.261	&	0.327	&		&		&	136	&	0.500	&	0.000	\\
	&		&	104	&	0.000	&	0.348	&		&		&	137	&	0.372	&	0.283	\\
Rh	&	45	&	103	&	0.055	&	0.289	&		&		&	138	&	3.546	&	0.225	\\
\\
\\
La	&	57	&	139	&	0.337	&	0.110	&	Lu	&	71	&	175	&	0.006	&	0.031	\\
Ce	&	58	&	140	&	0.894	&	0.089	&		&		&	176	&	0.002	&	0.000	\\
	&		&	142	&	0.000	&	0.115	&	Hf	&	72	&	176	&	0.008	&	0.000	\\
Pr	&	59	&	141	&	0.079	&	0.082	&		&		&	177	&	0.005	&	0.024	\\
Nd	&	60	&	142	&	0.227	&	0.000	&		&		&	178	&	0.021	&	0.022	\\
	&		&	143	&	0.037	&	0.065	&		&		&	179	&	0.007	&	0.015	\\
	&		&	144	&	0.105	&	0.094	&		&		&	180	&	0.035	&	0.020	\\
	&		&	145	&	0.020	&	0.049	&	Ta	&	73	&	181	&	0.009	&	0.013	\\
	&		&	146	&	0.091	&	0.053	&	W	&	74	&	182	&	0.024	&	0.012	\\
	&		&	148	&	0.004	&	0.044	&		&		&	183	&	0.013	&	0.007	\\
	&		&	150	&	0.000	&	0.047	&		&		&	184	&	0.029	&	0.013	\\
Sm	&	62	&	147	&	0.003	&	0.031	&		&		&	186	&	0.006	&	0.031	\\
	&		&	148	&	0.038	&	0.000	&	Re	&	75	&	185	&	0.004	&	0.014	\\
	&		&	149	&	0.005	&	0.031	&		&		&	187	&	0.001	&	0.033	\\
	&		&	150	&	0.022	&	0.000	&	Os	&	76	&	186	&	0.012	&	0.000	\\
	&		&	152	&	0.018	&	0.053	&		&		&	187	&	0.006	&	0.000	\\
	&		&	154	&	0.000	&	0.059	&		&		&	188	&	0.016	&	0.079	\\
Eu	&	63	&	151	&	0.000	&	0.042	&		&		&	189	&	0.004	&	0.111	\\
	&		&	153	&	0.002	&	0.048	&		&		&	190	&	0.021	&	0.168	\\
Gd	&	64	&	152	&	0.001	&	0.000	&		&		&	192	&	0.001	&	0.293	\\
	&		&	154	&	0.009	&	0.000	&	Ir	&	77	&	191	&	0.005	&	0.241	\\
	&		&	155	&	0.003	&	0.045	&		&		&	193	&	0.003	&	0.408	\\
	&		&	156	&	0.015	&	0.055	&	Pt	&	78	&	192	&	0.010	&	0.000	\\
	&		&	157	&	0.007	&	0.046	&		&		&	194	&	0.020	&	0.431	\\
	&		&	158	&	0.027	&	0.058	&		&		&	195	&	0.006	&	0.457	\\
	&		&	160	&	0.000	&	0.072	&		&		&	196	&	0.035	&	0.312	\\
Tb	&	65	&	159	&	0.004	&	0.060	&		&		&	198	&	0.000	&	0.099	\\
Dy	&	66	&	160	&	0.009	&	0.000	&	Au	&	79	&	197	&	0.010	&	0.176	\\
	&		&	161	&	0.004	&	0.075	&	Hg	&	80	&	198	&	0.035	&	0.000	\\
	&		&	162	&	0.016	&	0.101	&		&		&	199	&	0.016	&	0.043	\\
	&		&	163	&	0.002	&	0.093	&		&		&	200	&	0.051	&	0.030	\\
	&		&	164	&	0.018	&	0.091	&		&		&	201	&	0.020	&	0.027	\\
Ho	&	67	&	165	&	0.006	&	0.083	&		&		&	202	&	0.079	&	0.026	\\
Er	&	68	&	164	&	0.004	&	0.000	&		&		&	204	&	0.000	&	0.020	\\
	&		&	166	&	0.012	&	0.072	&	Tl	&	81	&	203	&	0.042	&	0.012	\\
	&		&	167	&	0.005	&	0.053	&		&		&	205	&	0.060	&	0.041	\\
	&		&	168	&	0.020	&	0.047	&	Pb	&	82	&	204	&	0.057	&	0.000	\\
	&		&	170	&	0.001	&	0.037	&		&		&	206	&	0.326	&	0.223	\\
Tm	&	69	&	169	&	0.006	&	0.031	&		&		&	207	&	0.313	&	0.280	\\
Yb	&	70	&	170	&	0.006	&	0.000	&		&		&	208	&	1.587	&	0.118	\\
	&		&	171	&	0.004	&	0.029	&	Bi	&	83	&	209	&	0.051	&	0.093	\\
	&		&	172	&	0.018	&	0.036	&	Th	&	90	&	232	&	0.000	&	0.042	\\
	&		&	173	&	0.008	&	0.031	&	U	&	92	&	235	&	0.000	&	0.006	\\
	&		&	174	&	0.040	&	0.037	&		&		&	238	&	0.000	&	0.020	\\
	&		&	176	&	0.000	&	0.030	&		&		&		&		&		\\

\enddata
\end{deluxetable}
\end{center}




\begin{thebibliography}{}

\bibitem[Cowan et al.(2002)]{cowan+02}
Cowan, J.\ J., et al.\ 
2002, \apj, 572, 861

\bibitem[Cowan et al.(2005)]{cowan+05}
Cowan, J.\ J., et al.\ 
2005, \apj, 627, 238

\bibitem[Cowan \& Sneden(2006)]{cs06}
Cowan, J. J., \& Sneden, C. 2006, Nature, in press

\bibitem[Cowan \& Thielemann(2004)]{ct04}
Cowan, J.\ J., \& Thielemann, F.-K., 2004, Phys.\ Today, 57, 47

\bibitem[Den Hartog et al.(2003)]{denhartog+03}
Den Hartog, E.\ A., Lawler, J.\ E., Sneden, C., \& Cowan, J.\ J.\ 2003, \apjs, 148, 543

\bibitem{den05}
Den Hartog, E. A.,  Herd, T. M., Lawler, J. E., Sneden, C.,
Cowan, J. J., \& Beers, T. C. 2005, \apj, in press

\bibitem[Lawler et al.(2006)]{den+06}
Den Hartog, E.\ A., Lawler, J. E., Sneden, C., \& Cowan, J.\ J.\ 2006, \apj,
to be submitted

\bibitem{iva06}
Ivans, I. I., \etal\ 2006, ApJ, in press 

\bibitem[K\"appeler, Beer \& Wisshak(1989)]{kbw89}
K\"appeler, F., Beer, H.\ \& Wisshak, K.\ 1989, Rep. Prog. Phys., 52, 945

\bibitem{kra06}
Kratz, K.-L., Farouqi, K., Pfeiffer, B., Truran, J. W., Sneden, C., \& 
Cowan, J. J.  
2006, 
ApJ, in preparation

\bibitem[Lawler et al.(2005)]{lawler+05}
Lawler, J.\ E., Den Hartog, E.\ A., Sneden, C., \& Cowan, J.\ J.\ 2005, \apj, in press

\bibitem[Lawler et al.(2004)]{lawler+04}
Lawler, J.\ E., Sneden, C., \& Cowan, J.\ J.\ 2004, \apj, 604, 850

\bibitem[]{lod03}
Lodders, K. 2003, \apj,  591, 1220

\bibitem[]{obr03}
O'Brien, S., et al. 2003, Pys. Rev. C, 68, 035801

\bibitem[]{pal00}
Palmeri, P., Quinet, P., Wyart, J.-F., {\&} Bi\'{e}mont, E. 2000, Physica
Scripta, 61, 323

\bibitem[Simmerer et al.(2004)]{simmerer+04}
Simmerer, J., Sneden, C., Cowan, J. J., Collier, J., Woolf, V. M., {\&}
Lawler, J. E. 2004, ApJ, 617, 1091

\bibitem[Sneden et al.(2003)]{Sneden+03}
Sneden, C., et al.  
2003, ApJ, 591, 936

\bibitem[Sneden \& Cowan (2003)]{Sne03}
Sneden, C., \& Cowan, J. J. 2003, Science, 299, 70 


\bibitem[]{wsgc}
Westin, J., Sneden, C., Gustafsson, B., \& Cowan, J.J. 2000, \apj, 530, 783

\bibitem[]{wIss} 
Wisshak, K., Voss, F., K\"appeler, F., Kazakov, L., \& Reffo, G. 1998, 
Phys. Rev. C, 57, 391

\end{thebibliography}
\end{document}